\documentclass[prl,aps,groupedaddress,floatfix,twocolumn,superscriptaddress]{revtex4}
\usepackage{amsmath,amssymb,multirow,epsfig,bm,pifont}
\usepackage[linkcolor=black,citecolor=black,urlcolor=blue,colorlinks=true,linktocpage=true]{hyperref}
\usepackage{graphicx}
\usepackage{epsfig}
\usepackage{bm}
\usepackage{tabularx}
\usepackage{makecell}

\usepackage{slashed}

\newcommand{\beq}{\begin{equation}}
\newcommand{\eeq}{\end{equation}}
\newcommand{\bea}{\begin{eqnarray}}
\newcommand{\eea}{\end{eqnarray}}
\newcommand{\tr}{\mathrm{tr}}

\newcommand{\bd}{\boldsymbol}

\begin{document}

\title{Tan's contact and the phase distribution of repulsive Fermi gases:\\ Insights from QCD noise analyses}

\author{William J. Porter}
\email{wjporter@live.unc.edu}
\affiliation{Department of Physics and Astronomy, University of North Carolina, Chapel Hill, NC, 27599, USA}

\author{Joaqu\'{\i}n E. Drut}
\email{drut@email.unc.edu}
\affiliation{Department of Physics and Astronomy, University of North Carolina, Chapel Hill, NC, 27599, USA}

\begin{abstract}
Path-integral analyses originally pioneered in the study of the complex-phase problem afflicting lattice calculations of finite-density quantum 
chromodynamics are generalized to non-relativistic Fermi gases with repulsive interactions. Using arguments similar to those previously applied 
to relativistic theories, we show that the analogous problem in nonrelativistic systems manifests itself naturally in Tan's contact 
as a nontrivial cancellation between terms with varied dependence on extensive thermodynamic quantities. We analyze that case 
under the assumption of gaussian phase distribution, which is supported by our Monte Carlo calculations and perturbative considerations.
We further generalize these results to observables other than the contact, as well as to polarized systems and systems with fixed particle number. 
Our results are quite general in that they apply to repulsive multi-component fermions, are independent of dimensionality or trapping potential, and 
hold in the ground state as well as at finite temperature.
\end{abstract}




\date{\today}
\maketitle

{\it Introduction.--}
Cold-atom experimentalists continue to engineer astonishing techniques to probe fundamental properties of quantum mechanics by means of multi-component gases~\cite{quantumGasMicroscope,entanglementGreiner,beamshapingGreiner,heidelberg2DExperiment,heidelbergBKTExperiment,ncsuShearViscosity,swinburne2DThermo,zwierleinPhaseSeparation,zwierleinSuperfluidity}.  
Access to the properties of these remarkable systems has expanded from the simplest thermodynamic quantities to observables characterizing 
nuanced dynamical and information-theoretic properties (see e.g.~\cite{dukeShockwaves,greinerETH}). Bridging a broad range of interactions, compositions, and dimensions, this 
ever-expanding repertoire of techniques is both celebrated and envied by theorists, as attempting to answer similar questions about low-temperature, strongly correlated fermions 
is accompanied by a long list of nontrivial complications~\cite{nishidaEpsilonExpansion,ContactReview}.

Of these impediments, one which is not only particularly formidable but is also shared with lattice studies of quantum chromodynamics (QCD) is the complex phase problem 
associated on the one hand with non-relativistic, repulsive or imbalanced Fermi systems and on the other with QCD at finite quark density~\cite{deForcrandSignProblem, signProblemGattringer}. In the QCD case, the
problem can be traced back to the breaking of time-reversal invariance at finite chemical potential (which also appears in quasi-relativistic systems like low-energy graphene away from the 
Dirac point)~\cite{GrapheneReview1,GrapheneReview2}, which bears a strong resemblance to the (mass- or spin-) imbalanced non-relativistic Fermi gas~\cite{parishPhaseDiagram}. Repulsive interactions, on the other hand, do not break time-reversal invariance per se, however that symmetry is lost upon decoupling via a Hubbard-Stratonovich transformation~\cite{stratonovich,hubbard}. As explained 
below, the partition function for $N_f$ identical fermion species then takes the path integral form
\beq
\mathcal{Z}^{}_{N^{}_{f}} = \int \mathcal D \phi\; {\det}^{N^{}_{f}}\mathcal{M}[\phi],
\eeq
where the matrix $\mathcal M$ and its determinant are generally complex, such that the latter cannot be used as a probability measure.
In all cases, such complex-valued fermion determinants lead to exponential cancellations creating unmanageable statistical uncertainties in Monte Carlo 
calculations~\cite{greensiteGaussianCorrections}. Because of its presence in the 
study of such a diverse class of physical systems, this difficulty has seen considerable attention from numerous perspectives ranging from polarized, low-dimensional systems, to 
ingenious density-of-states and complex Langevin approaches to studying lattice gauge theories~\cite{signProblemGattringer, AartsReview}. 
Sadly, as can be expected for such a ubiquitous affliction, no general solution to this problem is believed to exist~\cite{troyerSignProblem}.

Specific though these solutions must be, there is commonality between the remedies that do exist, and the sharing of intuition and techniques between the condensed-matter 
and high-energy communities has never failed to be a fruitful one. In Ref.~\cite{SplittorffTotalDerivative}, in particular, it was shown that 
by analyzing the distribution of the phase of the fermion determinant in finite-density QCD, it is possible to deduce the form of the associated free energies, and 
from them it is possible to cast sub-leading volume-dependent corrections to the baryon number in terms of a derivative. This insight allows for a 
clean demonstration of the origin of the associated signal-to-noise problem and also provides a general non-perturbative analytic tool for obtaining information about the 
distribution of this phase and in turn the behavior of the theory.

In this work, we generalize that noise analysis to nonrelativistic many-flavor fermions with repulsive interactions at finite temperature.  Our results hold for a much broader class of 
systems than discussed here, including all electronic systems. However, the simple example we provide is sufficient to demonstrate the techniques.  
For clarity, we maintain similar notation to Ref.~\cite{SplittorffTotalDerivative}, but we stress several points throughout our derivation as they differ significantly in our generalization of 
these techniques to polarized systems, which we present afterward.  Finally, after briefly describing lattice Monte Carlo calculations performed to justify some key assumptions, we 
detail the extension of these derivations to systems at fixed particle number.

{\it The phase distribution of repulsive Fermi gases.--}
In order to make contact with previous numerical work as well as that included in this manuscript, we perform our analysis in the Hamiltonian formulation beginning with 
a grand canonical Hamiltonian $\hat H - \mu \hat N$ with a zero-range interaction given by
\bea
\hat H \!-\! \mu \hat N \!=\! \int \! d^d r \Bigg[\sum_{s}^{}\hat{\psi}^{\dagger}_{s}(\bd r) K \hat{\psi}^{}_{s}(\bd r) +\frac{g}{2}\sum_{s\ne s'}^{}\hat n^{}_{s}(\bd r) \hat n^{}_{s'}(\bd r)\Bigg],
\eea
in terms of the differential operator $K = -\nabla^2/(2m) - \mu^{}_{}$, which incorporates the chemical potential $\mu$ and the flavor-$s$ fermion number density $\hat{n}^{}_{s} = \hat{\psi}^{\dagger}_{s}\hat{\psi}^{}_{s}$ which enters quadratically paired to the bare coupling $g$.

We place the theory on a discrete temporal lattice of dimensionless extent $N^{}_{\tau} = \left\lfloor\beta/\tau\right\rfloor\gg 1$, and after implementing a Trotter-Suzuki decomposition and an auxiliary field transformation (coupled to the density channel), we cast the partition function as
\beq
\label{eq:pathIntegralRepulsive}
\mathcal{Z}^{}_{N^{}_{f}} = \tr \left[e^{-\beta\left(\hat H - \mu \hat N\right)}\right] = 
\int \mathcal D \phi\;
{\det}^{N^{}_{f}}\mathcal{M}[\phi],
\eeq
%
%
%
The matrix $\mathcal M[\phi]$ encodes the dynamics of the system and separates into free and interacting components (see e.g. Ref.~\cite{MCMethods4} for further details):
\beq
\mathcal M[\phi] = \mathcal M^{}_{0} + A \delta\mathcal M[\phi]
\eeq
for sparse, block matrices $\mathcal M^{}_{0}$ and $\delta\mathcal M[\phi]$ and where
%
$A^2 = 2\left(e^{-\tau g} - 1\right)$.
%
While the coupling enters the integrand through the parameter $A$, the dependence of the full partition function 
on $g$ must necessarily be only through even powers of $A$, because we only have two-body interactions. This distinction is essential for our generalization.  For a repulsive interaction, $g>0$ implies that $A$ is purely imaginary so that under conjugation $A \mapsto -A$.  This sign reversal is analogous to the reversal of the sign of the chemical potential in this formalism's QCD application.

In the unpolarized case, we define the phase functional $\theta[\phi]$ for a given auxiliary field configuration $\phi$ through
\beq
\label{Eq:PolarDet}
{\det}^{}\mathcal{M}[\phi] = |{\det}^{}\mathcal{M}[\phi]| \;e^{i \theta[\phi]}.
\eeq
For the following analysis, we define the unquenched expectation value $\langle \cdot \rangle^{}_{N^{}_{f}}$ for a functional $X[\phi]$ as
\beq
\langle X \rangle^{}_{N^{}_{f}} = \frac{1}{\mathcal Z^{}_{N^{}_{f}}}\int \mathcal D \phi\;X[\phi]\;{\det}^{N^{}_{f}}\mathcal{M}[\phi]
\eeq
and the quenched expectation value for the same functional is
\beq
\langle X \rangle^{}_{|N^{}_{f}|} = \frac{1}{\mathcal Z^{}_{|N^{}_{f}|}}\int \mathcal D \phi\;X[\phi]\;|{\det}^{N^{}_{f}}\mathcal{M}[\phi]|
\eeq
with
\beq
\mathcal Z^{}_{|N^{}_{f}|} = \int \mathcal D \phi\;|{\det}^{N^{}_{f}}\mathcal{M}[\phi]|
\eeq

With these definitions, we express the probability density function for the phase in terms of the phase-quenched measure via
\beq
\label{Eq:ExpDelta}
\langle \delta(\theta - \theta^{}_{0})\rangle^{}_{N^{}_{f}} = e^{i N^{}_{f}\theta^{}_{0}} \frac{\mathcal Z ^{}_{|N^{}_{f}|}}{\mathcal Z ^{}_{N^{}_{f}}} \langle \delta(\theta - \theta^{}_{0})\rangle^{}_{|N^{}_{f}|},
\eeq
Representing the delta functions of Eq.~(\ref{Eq:ExpDelta}) in terms of their Fourier transforms, we are naturally led to consider
\beq
\label{eq:charUnqRepulsive}
\langle e^{i p \theta}\rangle ^{}_{N^{}_{f}} \propto \frac{1}{\mathcal Z^{}_{N^{}_{f}}}\left\langle\frac{{\det}^{p/2 + N^{}_{f}}\mathcal{M}[\phi]}{{\det}^{p/2}\mathcal{M}[\phi]^{*}}\right\rangle^{}_{}
\eeq
in the unquenched case and 
\beq
\label{eq:charQueRepulsive}
\langle e^{i p \theta}\rangle^{}_{|N^{}_{f}|} \propto \frac{1}{\mathcal Z^{}_{|N^{}_{f}|}}\left\langle\frac{{\det}^{p/2 + N^{}_{f}/2}\mathcal{M}[\phi]}{{\det}^{p/2-N^{}_{f}/2}\mathcal{M}[\phi]^{*}}\right\rangle^{}_{}
\eeq
in the quenched case for integer $p$, both of which can be verified using the polar form of Eq.~(\ref{Eq:PolarDet}). In both instances the proportionality constant is the normalization for the flat measure $\mathcal D \phi$ with the absence of a subscript indicating that the expectation is taken with respect to this measure.

As a result of: a) the peculiar combination of powers appearing inside the expectations present in Eqs.~(\ref{eq:charUnqRepulsive}) and~(\ref{eq:charQueRepulsive}); b) the properties of the matrix $\mathcal M[\phi]$ under conjugation; and c) the evenness of these expressions in the variable $A$; there exist transformations under which these expressions are invariant. This is in contrast to the analogous expressions in QCD, where conditions a) and b) provide the same invariance but when instead considered in combination with the nonzero quark chemical potential.  In particular, mapping $p\mapsto -2N^{}_{f} -p$ is equivalent to inverting the ratio in Eq.~(\ref{eq:charUnqRepulsive}).  This reversal combined with evenness in $A$ establishes the aforementioned invariance.

\begin{figure}[h]
\includegraphics[width=1.0\columnwidth]{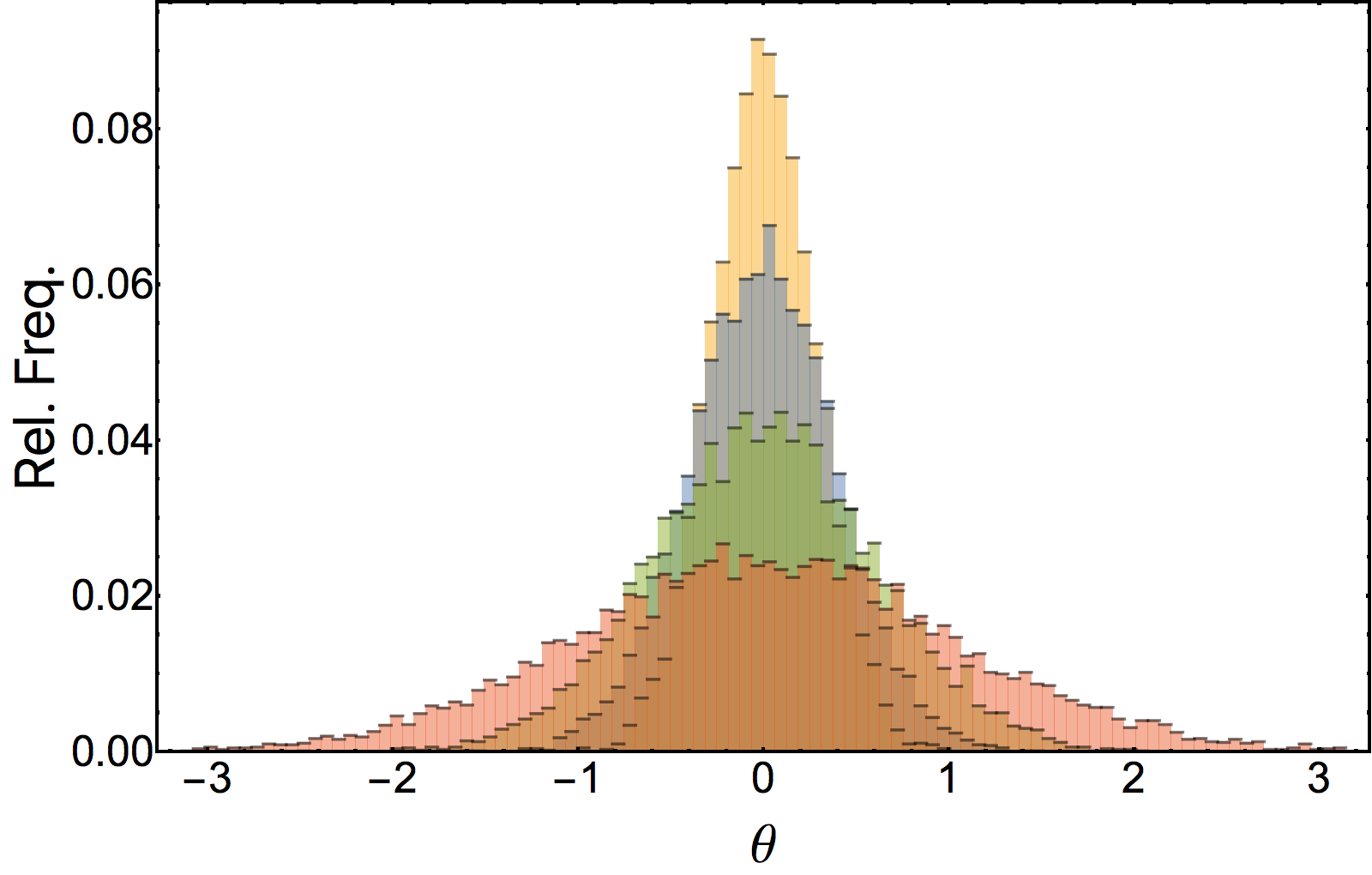}
\caption{\label{fig:histoUnpolarized}(color online) Distribution of the phase for an unpolarized system of nonrelativistic repulsive fermions in 1D at finite temperature for physical dimensionless coupling $\lambda = g \sqrt{\beta}$ of $\lambda = 0.25, 0.5, 1.0,$ and  $2.0$ (yellow, blue, green, and red, resp.); for chemical potential $\beta \mu \simeq -0.25$; and for lattice size $N^{}_{x} = 41.$}
\end{figure}

Although straightforward physical interpretation requires at least the restriction of $p$ to integer values, the functions of Eqs.~(\ref{eq:charUnqRepulsive}) and~(\ref{eq:charQueRepulsive}) are 
defined for arbitrary $p$.  The associated free energies $\ln\langle e^{i p \theta}\rangle ^{}_{N^{}_{f}}$ and $\ln \langle e^{i p \theta}\rangle ^{}_{|N^{}_{f}|}$ take 
the form of power series in $p$ with necessarily extensive coefficients encoding the remaining physics, as can be easily shown via the cumulant expansion.
Seeking a simple and convenient form for this power series, we note that for any $\delta p$, the transformation $p\mapsto \delta p - p$ is self-inversive so that for any function $f(p)$, the product 
$f(p) f(\delta p - p)$ is an invariant.  Taking $f(p) = p$ provides a parameterization of this series in terms of constants $X^{}_{j}$ for $0 \le j < \infty$ and powers of the monic quadratic polynomial 
$p(p+2 N^{}_{f})$.  Requiring the measure be normalized eliminates the $j = 0$ contribution, and we write
\beq
\label{eq:polyUnpolarized}
\ln \langle e^{i p \theta}\rangle ^{}_{N^{}_{f}} = -\sum_{j=1}^{\infty}X^{}_{j}\left[\frac{p}{2}\left(\frac{p}{2} + N^{}_{f}\right)\right]^{j}_{}.
\eeq
%

%
{\it Gaussian noise and Tan's contact.--}
It is desirable to truncate the series of Eq.~(\ref{eq:polyUnpolarized}) beyond the leading order in $p$ both for the sake of convenience and because such gaussian 
approximations are in many cases justified.  Support for this truncation in the context of QCD is provided by studies of the phase of the fermion determinant in lattice 
calculations \cite{qcdPhase}.  Although providing a characterization of the phase for each such system to which our generalizations are applicable is beyond the scope 
of this work, we do provide this inquiry for our prototype system.

Toward that end, we implement techniques similar to those in Ref.~\cite{MCMethods1,MCMethods2,MCMethods3,MCMethods4}, placing the system on a Euclidean spacetime 
of extent $N^{}_{x}\times N^{}_{\tau}$ periodic in the space and anti-periodic in (imaginary) time.  As shown in Fig.~\ref{fig:histoUnpolarized}, the phase assumes a roughly 
gaussian distribution over a broad range of couplings with a width that grows with increasing interaction strength as expected.
Approximate gaussian distributions were also found in Ref.~\cite{KaplanEtAl} for the logarithm of the fermion determinant in a sign-problem free case of
two-species fermions in the unitary limit, a property which was then used to predict the ground-state energy of the case of $N_f$ species~\cite{AmyPRL}. It is
remarkable that, in the complex-phase case, the phase angle (i.e. the imaginary part of the logarithm of the determinant) also displays a gaussian shape;
neither of these properties were expected, yet they are observed fairly universally (see e.g.~\cite{DeGrand, DrutPorterEE, qcdPhase}).

It is also worth noting that a perturbative expansion of Eqs.~(\ref{eq:charUnqRepulsive}) and~(\ref{eq:charQueRepulsive}) in powers of $A^2$ 
(odd terms do not contribute for pairwise interactions like the one considered here) reveals that calculating $X_j$ in Eq.~(\ref{eq:polyUnpolarized})
requires proceeding to order $j$ in $A^2$. In other words: at next-to-leading order, i.e. $A^2$, there are no contributions beyond $j=1$.
This suggests that the main qualitative features of the sign distribution can be captured perturbatively. Specific quantitative features can be expected to
be non-universal, however.

Hereafter, we truncate the expression in Eq.~(\ref{eq:polyUnpolarized}) beyond the first term, and following a Poisson resummation, the distribution of 
the phase takes the compactified gaussian form
\beq
\langle \delta(\theta - \theta^{}_{0})\rangle^{}_{N^{}_{f}} = e^{iN^{}_{f}\theta^{}_{0}+X^{}_{1}N^{2}_{f}/4}\frac{1}{\sqrt{\pi X^{}_{1}}}\sum_{k = -\infty}^{\infty} e^{-(\theta^{}_{0} +2\pi k)^{2}_{}/X^{}_{1}}.
\eeq

With this parameterization, we turn our attention to Tan's contact, which is governed by the on-site density-density correlation 
\beq
\hat{\mathcal C} = \frac{1}{2}\int d^{d}_{}r\; \sum_{s\ne s'}^{} \hat{n}^{}_{s}(\bd r)\hat{n}^{}_{s'}(\bd r),
\eeq
such that $\langle \hat{\mathcal C} \rangle_{N_f} = {\partial \ln \mathcal Z_{N_f}}/{\partial g}$.

Analyzing the phase-fixed quantity $\langle \delta(\theta - \theta^{}_{0})\hat{\mathcal C} \rangle^{}_{N^{}_{f}}$ by introducing a Fourier representation of
the delta function (as above), we note that
\beq
\label{Eq:ContactExp}
\langle  e^{ip\theta}\hat{\mathcal C} \rangle^{}_{N^{}_{f}}\!=\!
\frac{\partial}{\partial \bar g}\! 
\left \{
\frac{1}{\mathcal Z_{N_f}} 
\left\langle\frac{{\det}^{p/2}\mathcal{M}[\phi] {\det}^{N^{}_{f}}\mathcal{M}[\phi,g=\bar g]}{{\det}^{p/2}\mathcal{M}[\phi]^{*}}\right\rangle^{}_{} 
\!\right \}_{\bar g=g}.
\eeq

As above, we expect that the free energies of the expression in braces above take the form of 
polynomials in the variable $p$ with coefficients $c^{}_{k}$ that must depend on the coupling $\bar g$ and therefore
will be affected by the $\bar g$ derivative.
Using this form and identifying factors of $p$ with the application instead of derivatives $i \partial/\partial\theta^{}_{0}$, we obtain
\beq
\langle \delta(\theta - \theta^{}_{0})\hat{\mathcal C} \rangle^{}_{N^{}_{f}} = \left(c^{}_{0} +  i c^{}_{1}\frac{\partial}{\partial \theta^{}_{0}} - c^{}_{2}\frac{\partial^2}{\partial \theta^{2}_{0}} \dots\right)\langle \delta(\theta - \theta^{}_{0}) \rangle^{}_{N^{}_{f}},
\eeq
which, after integration over $\theta^{}_{0}$, provides the zero mode of the distribution, namely
\beq
\langle \hat{\mathcal C} \rangle^{}_{N^{}_{f}} = c^{}_{0},
\eeq
which can also be seen easily from Eq.~(\ref{Eq:ContactExp}) by setting $p=0$.

This result, as shown in Ref.~\cite{SplittorffTotalDerivative} for the baryon number in QCD, elucidates the nature of the sign problem in these systems:  The answer is entirely in the leading term; 
the subleading terms manifest the delicate cancellations that produce a reliable estimate of the observable.
This is in contrast to the result of applying the above derivative directly to our Gaussian form for the distribution of the phase and dropping terms that grow inversely in the moment $X^{}_{1}$.
Our generalization extends to statements made in the context of QCD regarding the orthogonality of the signal to the noise.

We have specialized the discussion to Tan's contact as thermodynamically conjugate to the coupling $g$, which is the natural generalization of the baryon number as conjugate to
the quark chemical potential as considered in Ref.~\cite{SplittorffTotalDerivative}. However, the result is in fact more general than advertised there: other one-body operators 
$\hat {\mathcal O}$ can be considered, with the corresponding modification of Eq.~(\ref{Eq:ContactExp}) when including such source terms $j \hat {\mathcal O}$. The differentiation 
with respect to $j$ would then be followed by the limit $j \to 0$, and the coefficients $c^{}_{k}$ are modified accordingly:
\beq
\langle \delta(\theta - \theta^{}_{0})\hat{\mathcal O} \rangle^{}_{N^{}_{f}}\! \!=\! 
\left(\!c^{}_{{\mathcal O}, 0} \!+\!  i c^{}_{{\mathcal O},1}\frac{\partial}{\partial \theta^{}_{0}} \!-\! c^{}_{{\mathcal O},2}\frac{\partial^2}{\partial \theta^{2}_{0}} \dots\!\right)
\!\!\langle \delta(\theta - \theta^{}_{0}) \rangle^{}_{N^{}_{f}}
\eeq
It is worth noting that the case of the contact is peculiar because, even though it is a two-body operator, the Hubbard-Stratonovich 
transformation allows one to compute it via a single derivative, as with the one-body operator described above.

{\it Polarized systems.--}
In order to investigate a polarized two-species gas~\cite{giorginiReview,revasym1,revasym2}, we return to the partition function given in Eq.~(\ref{eq:pathIntegralRepulsive}), taking the chemical potentials $\mu^{}_{\downarrow} \ne \mu^{}_{\uparrow}$ to be distinct and writing
\bea
\label{eq:pathIntegralRepulsivePolarized}
\mathcal{Z}^{}_{\uparrow\downarrow}& = & \tr \left[e^{-\beta\left(\hat H - \mu^{}_{\uparrow} \hat{N}^{}_{\uparrow} - \mu^{}_{\downarrow} \hat{N}^{}_{\downarrow}\right)}\right] \\
& = & \int \mathcal D \phi\;{\det}^{}\mathcal{M}^{}_{\uparrow}[\phi]\,{\det}^{}\mathcal{M}^{}_{\downarrow}[\phi],
\eea
where we have added an additional indication of which chemical potential appears in the fermion matrix $\mathcal{M}^{}_{\uparrow,\downarrow}[\phi]$. We then write
\beq
{\det}^{}\mathcal{M}^{}_{s}[\phi] = |{\det}^{}\mathcal{M}^{}_{s}[\phi]| \;e^{i \theta^{}_{s}[\phi]}
\eeq
for $s = \uparrow,\downarrow$.  In a fashion analogous to that presented previously, considering the joint distribution $\langle \delta(\theta^{}_{\uparrow} - \theta^{}_{\uparrow, 0}) \delta(\theta^{}_{\downarrow} - \theta^{}_{\downarrow, 0}) \rangle^{}_{\uparrow\downarrow}$ immediatly yields a relationship between this distribution and its value relative to the phase-quenched measure.  As before, this naturally motivates the investigation of joint characteristic functions of the form 
\beq
\langle e^{i p \theta^{}_{\uparrow}}e^{i q \theta^{}_{\downarrow}}\rangle ^{}_{\uparrow\downarrow} \propto \frac{1}{\mathcal{Z}^{}_{\uparrow\downarrow}}\left\langle\frac{{\det}^{(p+2)/2}\mathcal{M}^{}_{\uparrow}[\phi]}{{\det}^{p/2}\mathcal{M}^{}_{\uparrow}[\phi]^{*}}\frac{{\det}^{(q+2)/2}\mathcal{M}^{}_{\downarrow}[\phi]}{{\det}^{q/2}\mathcal{M}^{}_{\downarrow}[\phi]^{*}}\right\rangle^{}_{}
\eeq
by means of their symmetries.  Careful examination precludes a parameterization as simple as the one given in Eq.~(\ref{eq:polyUnpolarized}): the characteristic functions must be invariant under the transformations $p\mapsto -p-2$ and $q\mapsto -q-2$ taken together, but separately these replacements are not permitted.  This condition can be ensured by taking
\bea
\label{eq:polyPolarized}
\ln \langle e^{i p \theta^{}_{\uparrow}}e^{i q \theta^{}_{\downarrow}}\rangle ^{}_{\uparrow\downarrow} = -\sum_{j=1}^{\infty} C^{}_{j}(p,q).
\eea
where
\beq
\label{eq:summandPolarized}
C^{}_{j}(p,q) = A^{\uparrow}_{j}p^{j}_{}(p + 2)^{j} + A^{\downarrow}_{j}q^{j}_{}(q + 2)^{j} - B^{}_{j}(p-q)^{2j},
\eeq
and where the moments $A^{\uparrow}_{j}, A^{\downarrow}_{j},$ and $B^{}_{j}$ depend on both chemical potentials. This requirement can be seen in the residual dependence of the measure on say $\mu^{}_{\uparrow}$ even in the case where $p = 0$ and similarly for the case where $q = 0$.  More information can be gleaned immediatly by noting that we may exchange the chemical potentials if we similarly perform the swap $p \leftrightarrow q$.  This observation relates the coefficients for the homogeneous terms.  Finally, it is straightforward to relate the diagonal of these coefficients to the moments $X^{}_{j}$ by taking the chemical potentials to be as in the previous section and equating the phases.  Truncating this expression again at first order, we may again perform a Poisson resummation, and the result of this calculation provides access to an analysis similar to that previously obtained for the contact.
More general systems including copies of each flavor are approachable by essentially the same techniques~\footnote{In that case, Eq.~(\ref{eq:summandPolarized}) is modified in two ways.  First, the homogeneous terms are invariant under different transformations, each depending on the relative abundance of each flavor.  Second, the factors comprising the inhomogeneous term also include as a summand the difference between these two transformations so that it remains invariant under their combination.}.

{\it Finite systems.--}
In order to comment on systems at fixed particle number, we restrict the grand-canonical partition function via Fourier projection writing the canonical partition function for an $N$-particle system via
\beq
Q^{}_{N} = \int \mathcal D \phi \; P^{}_{N}[\phi]
\eeq
where 
\beq
P^{}_{N}[\phi] = \frac{1}{2\pi}\int_{0}^{2\pi}\!\! d\alpha\;e^{-i N \alpha}\,{\det}^{N^{}_{f}}_{}\!\left(\openone + e^{i \alpha}\mathcal{U}[\phi]\right),
\eeq
where the matrix $\mathcal{U}$ contains all the physical input for the system (see e.g.~\cite{MCMethods4}).
Analysis of this measure leads to characteristic functions of the form
\beq
\frac{1}{Q^{}_{N}}\int \mathcal D \phi \; P^{}_{N}[\phi]\,e^{i p \theta } \propto \frac{1}{Q^{}_{N}}\left\langle\frac{P^{(p+2)/2}_{N}[\phi]}{P^{p/2}_{N}[\phi]^{*}_{}}\right\rangle,
\eeq
where now we must require integer $N$ so that by a change of variables, the denominator can be rewritten so as to be accessible by techniques described earlier.  After changing variables, we find again that the conjugation amounts to reversing the sign of $A$, and the above expressions are invariant under the transformation $p \mapsto -p-2$.  The analysis then proceeds as detailed previously.

{\it Summary and Conclusions.--}
In this work, we have elucidated the origin of the debilitating fluctuations in lattice Monte Carlo calculations of the Tan contact in repulsive Fermi systems, and we investigated similar issues in polarized nonrelativistic gases as well as in systems projected to fixed particle content. We have, furthermore, generalized the analysis of Tan's contact to arbitrary one-body operators.

We accomplished the above by generalizing techniques used to study the baryon number for finite-density QCD, and by showing that similar arguments provide insight into the complex phase 
problem encountered in the exploration of the repulsive sector of the parameter space.  In these systems, some previous analytic work carries over almost without modification, although key 
dissimilarities force a different correspondence: the phase problem in finite-density QCD is due to explicit breaking of time-reversal invariance, while in repulsive Fermi gases it is due to 
the signature of the interaction; furthermore, rather than gleaning information about particle number from the fluctuations in the action, we are led naturally to relations involving Tan's contact.

We perform lattice Monte Carlo calculations to verify that the onset of these phase problems indeed present quasi-Gaussian phase distributions.  Studying repulsive, finite-temperature 
fermions in one spatial dimension, we found results similar to those originally used to justify the application of these techniques in relativistic theories. A perturbative analysis of the problem 
indicates a direct connection between moments of the phase distribution and orders in perturbation theory: at a given order in the latter, contributions are only present in moments of up to a 
fixed order and otherwise vanish.

After providing this analogy and detailing the numerical calculations required to verify key assumptions, we demonstrated that the technique applies further still to the case of a general 
two-species gas, and we have detailed the modifications made to pivotal algebraic arguments. Polarized systems require that the generally distinct fermion determinants be treated 
independently, and the two resulting phase angles lead to generalized free energies that are multi-variate polynomials.  Finally, we show that our insights apply equally well to systems at 
fixed particle number.

Although we have considered the case of two-body contact interactions exclusively, we expect our results to generalize to more general interactions and to mixed Bose-Fermi 
ensembles, a subject of growing relevance~\cite{BFMix1,BFMix2,BFMix3,BFMix4,BFMix5}.  We leave such investigation for future study.

\acknowledgments
This material is based upon work supported by the 
National Science Foundation under Grants No. 
PHY{1306520} (Nuclear Theory Program), and
PHY{1452635} (Computational Physics Program).



\end{document}